# Mathematical model of COVID-19 intervention scenarios for São Paulo- Brazil

Osmar Pinto Neto[1,2,4*], José Clark Reis[2], Ana Carolina Brisola Brizzi[1,2], Gustavo José Zambrano[2], Joabe Marcos de Souza[2,3], Wellington Amorim Pedroso[1,2], Rodrigo Cunha de Mello Pedreiro[1,6,7], Bruno de Matos Brizzi[2], Ellysson Oliveira Abinader[5], Deanna M. Kennedy[8], Renato Amaro Zângaro[1,4].

[1] Anhembi Morumbi University, Biomedical Engineering Department, São Paulo, SP, Brazil.

[2] Arena235 Research Lab – São José dos Campos, SP, Brazil.

[3] Universidade de São Paulo, Departamento de Engenharia Aeronáutica, São Paulo, SP, Brazil.

[4] Center for Innovation, Technology and Education – CITE, Parque Tecnológico de São José dos Campos, São José dos Campos, SP, Brazil.

[5] Instituto Abinader, Manaus , AM, Brazil.

[6] Estácio de Sá University, Nova Fribugo, RJ, Brazil.

[7] Santo Antônio de Pádua College, Santo Antônio de Pádua, RJ, Brasil.

[8] Texas A&M University, TX, USA

Address Corespondence to:

Osmar Pinto Neto

osmarpintoneto@hotmail.com

Parque Tecnológico de São José dos Campos, Estrada Dr. Altino Bondensan, 500, São José

dos Campos, SP, Brazil, 12247-016.

**05/17/2020**

**Abstract**

An epidemiological compartmental model was used to simulate social distancing strategies to contain the COVID-19 pandemic and prevent a second wave in São Paulo, Brazil. Optimization using genetic algorithm was used to determine the optimal solutions. Our results suggest the best-case strategy for São Paulo is to maintain or increase the current magnitude of social distancing for at least 60 more days and increase the current levels of personal protection behaviors by a minimum of 10% (e.g., wearing facemasks, proper hand hygiene and avoid agglomeration). Followed by a long-term oscillatory level of social distancing with a stepping-down approach every 80 days over a period of two years with continued protective behavior.

Keywords: COVID-19, mathematical modeling, compartmental model, social distancing, pandemic.

# Mathematical model of COVID-19 intervention scenarios for São Paulo- Brazil

The World Health Organization (WHO) officially declared COVID-19 a pandemic in March of 2020[1]. Two months later over 4.3 million people have tested positive for the virus resulting in more than 300,000 deaths worldwide[2]. Most countries, including Brazil, have implemented widespread social distancing (SD) restrictions in an effort to mitigate the spread of COVID-19. It appears, from updated models, that these strategies have effectively reduced the number of cases and associated deaths compared to earlier predictions[3]. However, a new report indicated that SD measures in Brazil have not been as effective at reducing the reproduction number ($R_0$) as it has in other countries[4]. $R_0$ represents the average number of secondary cases that result from the introduction of a single infectious case in a susceptible population[5]. The $R_0$ for Brazil is 2.81, while the $R_0$ for China, for example, dropped to below 1 with SD measures in place[6]. A $R_0 = 2.81$ indicates that Brazil has not controlled the COVID-19 pandemic. In fact, Brazil is the new virus hot spot with the highest infection and death counts in Latin America, with the State of São Paulo considered one of the main hotspot[6]. São Paulo is the most poulous state in Brazil with over 46 million inhabitants. Therefore, exploring alternative SD intervention strategies for Brazil in general and São Paulo specifically is of utmost importance.

Several factors may lead to differences in intervention strategies on COVID-19 infection and death rates for São Paulo, Brazil compared to other states and countries. For example, research has indicated the older adults (60+) are at the greatest risk of experiencing complications from COVID-19[7]. Given that Brazil has a large percentage of its population over the age of 60 years, especially in urban areas such as São Paulo, the incidence of aggravated cases may be particularly high in comparison to places with younger population demographics[8]. It has also been debated whether environmental conditions influence the behavior of the COVID-19 virus similar to the common cold and flu[9,10]. Yet, the majority of the research investigating the virus has been conducted in environments different/opposite than that of Brazil and other areas in the southern hemisphere. In addition, Brazil faces many economical and sociocultural challenges that may affect mitigation strategies differently than cities, states, and/ or countries that are the current focus of most prediction models[8]. Therefore, the purpose of the current investigation is to model COVID-19 SD intervention strategies on transmission dynamics in São Paulo, Brazil and to determine best-case scenarios.

**Results**

Our model was able to accurately fit the corrected accumulated cases and deaths data for Brazil and São Paulo (Figure 1). On the day of the analysis (May 11, 2020), SD was estimated at 41% in Brazil and 52% in São Paulo and protection was 59% in Brazil and 60% in São Paulo. All optimized coefficients, as well as the latent period, infectious period, hospitalized period, ICU period estimated from the model can be found in Table 1.

Considering the size of the country and that many regions are facing different stages of the pandemic, we concentrate further analyses in the state of São Paulo. The results suggest that the optimal strategy was the stepping down strategy (Figure 2a,b) and optimal time window was 80 days (Figure 2c). The results suggest that by constraining SD and protection levels to realistic ranges (30 to 70% and 50 to 70% respectively), optimal solutions regarding strategies and time window to contain the first and second peak of the pandemic converge confirming the optimal strategy (Figure 2b) and window (Figure 2c) that can be observed in (Figure 2a)

*Optimal solution for São Paulo*

Considering keeping the current SD and protection values and the optimal strategy and time window it would be possible to contain the first peak in the pandemic but not the second (Figure 1e); unless, at least the current SD were kept indefinitely. Considering scenarios where SD would eventually either drop or oscillate, optimal strategy and optimal time window suggest that critical cases over ICU threshold would be approximately 20 thousand; at 80-day windows stepping strategy would cause approximately a 90 thousand decrease in the number of critical cases over ICU threshold in comparison to the intermittent strategy and 90 thousand compared to the constant SD strategy. Additionally, considering the stepping strategy, an 80-day window would cause approximately 70 thousand decrease in the number of critical cases over ICU threshold compared to a 40-day window and 30 thousand compared to a 60-day window. Our results indicate that keeping protection as it is now, a lock-down scenario of SD=75% for the next 60 days followed, by a 40-day window lock-down stepping strategy, would contain the first peak of the pandemic and cause a significant drop in the total number of cases for the till October 2020; however, if protection level drops with time it may have cause a second large peak.

Nevertheless, the results suggest that by keeping current SD, increasing protection by 10%, and using optimal strategy and time windows between 60-80 days, it is possible to contain the first peak and second peaks of the pandemic in São Paulo (Figure 1f). After a 60 days period, a reduction in average SD across the reminder of the pandemic with 60-80-day windows steeping strategy would cause approximately a 30 thousand decrease in the number of critical cases over ICU threshold in comparison to the 80-day window intermittent strategy and 100 thousand compared to the constant SD strategy. Additionally, considering the stepping strategy, either a 60-80-day window would cause approximately 40 thousand decrease in the number of critical cases over ICU threshold compared to a 40-day window.

Finally, considering increasing protection by 10% but having a drop in SD from 52% to 40%, optimal strategy and optimal time window suggest that critical cases over ICU threshold would be approximately 30 thousand; at 80-day windows stepping strategy would cause approximately a 11 thousand decrease in the number of critical cases over ICU threshold in comparison to the intermittent strategy and 100 thousand compared to the constant SD strategy (Figure 1g). Additionally, considering the stepping strategy, an 80-day window would cause approximately 50 thousand decrease in the number of critical cases over ICU threshold compared to a 40-day window and 15 thousand compared to a 60-day window.

**Discussion**

We used a SUEIHCDR compartmental model to project thousands of scenarios for the transmission dynamics of COVID-19 in São Paulo, Brazil through the next two years. We used advanced algorithms to model scenarios related to strategy type, SD magnitude, time window, and level of personal protection. The goal was to determine the best-case scenario to control the current peak of infections and avoid a second pandemic wave.

*Controlling first peak*

Currently, Brazil has the highest rate of transmission in the world with an estimated $R_0$ of 2.81[6], indicating that it has yet to contain the first peak in infections and associated deaths due to the COVID-19 pandemic. Data from around the world (e.g., Asia, Europe, and North America) indicate that it is possible to mitigate the spread of COVID-19 with widespread SD and PPM measures[3]. However, our analysis of location data[11,12] indicate that the current level of SD is at only 42% for Brazil and for 52% São Paulo. In addition, the current protection values are 60% for the state and 69% for the country. Not only does our model indicate that current SD and protection values are insufficient in controlling the pandemic, they will have dire consequences on the overall number of infections and associated deaths with an extremely large first peak. In addition, our model suggest that this will result in the need of public health resources, especially ICU, exceeding what is currently available. With the current levels of SD and protection, our model predicts that the ICU needed will surpass the available ICU in Sao Paulo at the first drop of SD levels, unless protection increases (Figure 1). The scenario may be even worse because not all ICU beds available are exclusively dedicated to covid-19 patients and because some cities may experience the health-care system failure before others.

According to our model, it is possible for the state of São Paulo to gain control over the first peak if levels of personal protection are significantly increases (at least 10%). Alternatively, an immediately increase SD to values over 75% may solve the problem of the first peak as well. However, increasing SD to values in 75% may result in a second peak if the restrictions are lifted within 40days. SD at 75% represents a complete lock down[13]. Thus, such a strategy should be used with caution. Further, it should only be used in critical cities and not the State as a whole. Our model suggest that if this high-level of SD is used, it is only necessary for a short time (until end of June 2020).

Note, however, that lower levels of SD will need to be in place for years to come to maintain control over the pandemic[14,15]. Widespread use of PPM (e.g., wearing facemasks, frequently washing hands, using hand sanitizer, maintaining physical distance between other people, and avoiding agglomerations) and high rates of testing have been emphasized to mitigating the spread of COVID-19 in addition to SD[15-19]. The results of our model agree; our model indicates that an increase in personal protection to a level of 70% for São Paulo in combination with SD is necessary to contain the concurrent peak in infections and associated deaths. Eikenberry et al. (2020) estimated that the efficacy of using face masks ranges between 50 to 90%, depending upon mask material and fitting[20]. They assumed that at least 50% protection factor would be achievable for well-made and well-fitted mask usage by the entire population[20]. The fitting of our model predicts that current level of protection is about 60%. An increase in protection levels would need a massive effort by public health officials to enforce and/or educate people to use facemask and to maintain a 2 meters safe distance from other people. Note, some Brazilian cities have implemented strict SD guidelines and require some level of personal protection[19]. For example, Belo Horizonte, Rio de Janeiro, and Salvador require facemasks in public[21]. São Paulo recently announced they will require facemask as well.

*Avoiding a second peak*

Assuming that Brazil can effectively reduce $R_0$ below 1 with SD and personal protection, it is important to take steps to reduce the likelihood of a second peak. Our models indicate that if SD and personal protection measures are stopped too soon or reduced too much after the containment of the first peak, a second peak in infections and associated deaths will occur. Many experts agree that a secondary pandemic wave is likely if SD restrictions are lifted too quickly[22,23]. Therefore, determining when and how to relax restrictions has become the focus of epidemiological work around the world. It has been proposed that a responsible exit strategy should continue SD restrictions alongside widespread testing and contract tracing[24]. Laboratory testing to confirm COVID-19 exposure and/or diagnosis has been a major obstacle for the mitigation efforts worldwide. Unfortunately, testing efforts in Brazil have been deficient. Brazil has conducted the least COVID-19 tests per capita worldwide (https://www.worldbank.org/). In addition, reports indicate that Brazil uses substandard testing kits and only tracks hospitalized cases (https://p.dw.com/p/3cBQi). The lack of data for contact tracing in Brazil suggest efforts are insufficient as well.

Nevertheless, our models suggest it is possible for Brazil, considering São Paulo as a model state, to avoid a second peak. According to our results, the best-case exit strategy to prevent a second peak in São Paulo was a stepping-down strategy over a two-year period. A stepping-down approach would involve a gradual stair-step down. For example, the stepping-down approach we modeled multiplied SD values of 40%, 30% and 20% by 1, then 1/2, then 1/4, then 1/8 and then back to 1 and the we repeated each stair-step down. A stepping-down strategy was also the best-case exit strategy modeled for the US[15]. A stepping-down approach may be beneficial because it allows for periods of transmission leading to heard immunity without overwhelming public health resources[14,25]. Alternatively, a one-time SD may result in a catastrophic second peak (Figure 1, yellow curve), if the virus reoccurred and not enough people have immunity[14,25]. In addition, the stepping-down approach resulted in a 6.5% reduction in total time required to SD over the two-year period, potentially reducing the economic and social costs associated with SD.

The results of the current investigation suggest that an 80-day window between each step was the most beneficial strategy. This result is also consistent with that for the US[15]. Note, however, the current investigation modeled more time windows and more complex algorithms than the US model. In addition, the best-case stepping-down strategy with an 80-day windows was a magnitude of SD of at least 50%, in the highest windows, or approximately 24% average through the period of analysis. Furthermore, the results suggest that higher protective measures could account for lower SD values, which may be associated with economic and/or psychological benefits. Alternatively, higher SD values delay the onset of the second peak. Delaying the onset of the peak may allow Brazil to procure additional public resources and more time for the development of effective treatments or even a vaccine.

The minimum protection rate for this scenario was 70% for São Paulo. If protection rates are maintained at these levels the models suggest that a second peak would be avoided. However, if protection drops, a second peak would occur and would cross the threshold for available public resources with perhaps devastating consequences. Mortality rates associated with COVID-19 may rise when hospitals become overwhelmed and have fewer resources to treat patients with life-threatening symptoms (https://www.hopkinsmedicine.org/coronavirus). This result provides additional support for the notion that personal protection is critical for maintaining control over the COVID-19 pandemic[15-19].

**Methodology**

*The SUEIHCDR model*

We extended a generalized Susceptible-Exposed-Infected-Recovered (SEIR) compartmental model using factors specific to COVID-19[26,27] to investigate the COVID-19 pandemic in the US. It is composed of eight compartments: Susceptible, Unsusceptible, Exposed, Infected, Hospitalized, Critical, Dead, and Recovered (SUEIHCDR) (Figure 1).

The SUEIHCDR model assumes that as time progresses, a susceptible person out of the population ($N_{pop}$) (Equation 1) can either become unsusceptible (Equation 2) considering a protection rate ($α$; Equation 9; [23]) or exposed (Equation 3) to the virus considering social distancing ($SD$) and an infection rate ($β$). This protection rate was introduced to account for possible decreases in the number of susceptible people to the virus caused by factors other than social distancing such as the use of facemasks and hand washing. Exposed people become infectious after an incubation time of $1/γ$ (Equation 4). Infected people stay infected for a period of $1/δ$ and can recover with no medical attention ($m$) or can be hospitalized ($1-m$). Hospitalized people (Equation 5) stay hospitalized for $1/ζ$ days and can either recover ($1-c$) or become critical ($c$) needing to go to an Intensive care unit (ICU). A person stays on average $1/ε$ in the ICU (Equation 6) and can either go back to the hospital ($1-f$) or die ($f$; Equation 7). Recovered people (Equation 8) can come from infection ($m$) or from the hospital ($1-c$).

$$\frac{dS(t)}{dt} = -\frac{(1-SD(t))βS(t)I(t)}{N_{pop}} - α(t)S(t) \tag{1}$$

$$\frac{dU(t)}{dt} = α(t)S(t) \tag{2}$$

$$\frac{dE(t)}{dt} = +\frac{(1-SD(t))βS(t)I(t)}{N_{pop}} - γE(t) \tag{3}$$

$$\frac{dI(t)}{dt} = +γE(t) - δI(t) \tag{4}$$

$$\frac{dH(t)}{dt} = +(1-m)δI(t) + (1-f)εC(t) - ζH(t) \tag{5}$$

$$\frac{dC(t)}{dt} = +c\zeta H(t) - \varepsilon C(t) \tag{6}$$

$$\frac{dD(t)}{dt} = f\varepsilon C(t) \tag{7}$$

$$\frac{dR(t)}{dt} = +m\delta I(t) + (1-c)\zeta H(t) \tag{8}$$

$$\alpha(t) = \alpha_0 \frac{\log(t+1)}{\log(t_f)} \tag{9}$$

where $\alpha_0$ is the maximum or minimum possible value for a window of time and $t_f$ is the final time of the window. Alpha was optimized considering the window of time from the beginning of the pandemic until present day, and it was manipulated afterwards in different windows of time to project possible future scenarios. Changes in α causes changes in the number of people that are unsusceptible to the disease at a given time; we exhibit alpha manipulations as the percentage ratio of the unsusceptible or protected people over the country's population (i.e. Protection (%)). Note, that insusceptibility in our model accounts for the time-dependent state of an individual or behavior that can take someone from susceptibility to insusceptibility or the other way around.

Furthermore, social distancing until present day was determined from mobility trends data from Apple Maps[11] and community mobility reports from Google[12]. Data were low-pass filter filtered at 0.09 Hz (Butterworth 4th order), and percentage changes from baseline were considered (Apple's driving data was averaged with Google's retail and recreation, grocery and pharmacy, parks, transit stations, and workplaces average percent change from baseline. From present day forward *SD* was manipulated in different windows of time to project possible future scenarios.

*Solving and testing the model*

We used the fourth order Runge-Kutta numerical method to solve our system of ordinary differential equation in MATLAB (MathWorks Inc.R14a).

*Model's Coefficients Optimization*

We used both accumulated case and deaths time series to fit the model, both corrected by a factor (Table 1). Cases sub-test factor was calculated as the ratio between the death rate in Iceland (country with the greatest percentage of test per inhabitant[28]), corrected by age stratification (older adults (60+): 6.4%; senior older adults (80+): 13.4%[29]). Death sub-factor was determined comparing data from deaths in the last months to average deaths in the same period from past 5 years[30-32]. The resulting sub-factor was similar to that reported by other sources (https://www.ft.com/content/6bd88b7d-3386-4543-b2e9-0d5c6fac846c).

Fitting analysis was done with a custom build MATLAB global optimization algorithm using Monte Carlo iterations and multiple local minima searches (Figure 1). The algorithm was tested for the best solution considering all inputs varying within ranges obtained from the WHO[33] and several publications[34-36] (Table 1). Infected initial values ($I0$) were determined from corrected accumulated cases as well as initial death values ($D0$). Other initial values were set proportional to $I0$ considering model parameters ($m, c, f$); all initial parameters could vary during optimization as well (Table 1). Data's start date was Feb 25, 2020, and end date was Dec 25, 2021. Data under 500 active cases were discarded. Data under 500 active cases were discarded. A Confidence Interval of 95% was estimated using Monte Carlo for a 2% error for all coefficients used to evaluate the confidence in the model results, to test if future projected scenarios were statistically different at a 5% level, and to compare model results to actual ICU numbers (Figure 1). Note confidence interval projected lines were not included in all the figures for clarity. Model results were based on an average of 5000 runs. Peak ICU estimations were compared to data obtained from DATAUSA coronavirus database (https://datausa.io/coronavirus).

*Future Projections*

After model's coefficient were fitted an optimization workflow using ESTECO's mode Frontier (Esteco s.p.a; 2017R4-5.6.0.1) was implemented to find the optimal mitigation strategy out of the proposed here could be identified. Future scenarios considered changes in SD and protection staring June 1st, 2020. First, we run a DOE (Design of Experiment) scheme of around 1000 SOBOL individuals and 1000 Latin Hypercube individuals created by varying maximum SD values per window within 0-75% and protection percentage between 20-95%. Then data was constrained considering more realistic ranges of SD and protection (30 to 70% and 50 to 70%, respectively) (complete optimization analysis can be found in the supplementary file). We propose 3 strategies of mitigation (Figure 2): 1) a stepping- down strategy (starting at a specific SD, it is divided by half for the next 3 time windows, on the fourth time window SD is back to its initial value, and the process is repeated); 2) a standard intermittent SD strategy (a specific SD value alternates with periods of no SD); and 3) a constant SD strategy (SD is kept constant at as specific value); with each strategy considering three different times windows: 40, 60 or 80 days (Figure 3). Furthermore, strategies were compared using similar average SD across time; note however, that for because of its design when intermittent and stepping down strategies have the same maximum SD, and when adopting half of this values for the constant SD strategy, average SD values across time tends to be 6.25% smaller for the stepping-down strategy compared to the other two. A MOGA - Multi Objective Genetic Algorithm was used to drive the optimization, due to the discrete nature of the Variables Strategy and Window. We minimize the number of critical cases over ICU threshold (ICU_E) for the duration of the analysis, the number of critical cases over ICU threshold in the first peak of the pandemic (ICUE_1), the number of critical cases over ICU threshold in the second peak of the pandemic (ICUE_2) and SD.

**Acknowledgements**

The authors would like to thank Apple, CDC, IHME, Google and journals that have made data and information relative to COVID-19 publicly available. We would also like to thank all the frontline workers risking their lives to help others during the COVID-19 pandemic.

**Author contribution**

Osmar Pinto Neto, José Clark Reis and Renato Amaro Zângaro conceived and planned the study; Osmar Pinto Neto, Gustavo José Zambrano and Joabe Marcos de Souza developed the mathematical model. Wellington Pedroso E. Amorim, Rodrigo Cunha de Mello Pedreiro, Bruno de Matos Brizzi and Ellysson Oliveira Abinader worked on revising previous studies and data gathering. Osmar Pinto Neto designed the computational framework and performed the computations. José Clark Reis, Ana Carolina Brisola Brizzi and Deanna M. Kennedy wrote the introduction. Osmar Pinto Neto and Joabe Marcos de Souza wrote the methods and results sections. Osmar Pinto Neto, Deanna M. Kennedy and Gustavo José Zambrano wrote the discussion. All authors made substantial contributions to the conception and design of the work. All authors have discussed, revised and approved the contents of the final manuscript. All authors have agreed to be accountable for all aspects of the work in ensuring that questions related to the accuracy or integrity of any part of the work are appropriately investigated and resolved.

**Competing interest statement**

The authors declare no conflict of interest.

**Table Legends:**

Table 1: Optimized coefficients for Brazil (BR) and the state of São Paulo (SP) on May 11, 2020, respective ranges and correcting factors for accumulated cases (Factor C) and deaths (Factor D).

**Figure Legends:**

Figure 1: Fitting results for corrected accumulated cases and deaths Brazil (a,b) and for São Paulo (c,d) with a 95% confidence interval and representative results for intensity care units bed occupancy per day (ICU_PD; i.e. critical cases). e) shows 8 different cases with protection kept at current state level of 60%.; for the first 6 scenarios SD is kept at current levels (SD=52%) during the high SD window; scenarios 7 and 8, SD is constant at 26% and 52%. Finally, the last scenario (yellow) shows a 75% SD (lockdown) with stepping strategy with 40-day windows; f) show similar scenarios but with 70% protection; g) show 70% protection and 40% SD; h) show changes in SD through time.

Figure 2: Optimization results for state of São Paulo showing the influence of strategy (a,b) and window size (c) on the total number of critical cases over ICU threshold for the whole period of analysis (end day Dec 25, 2021; ICU_E) and on the number of critical cases over ICU threshold for the first (ICU_E1) and second peak of the pandemic (ICU_E2). Red color indicate either constant SD strategy (a,b) or 80-day windows (c); green color indicates either intermittent strategy (a,b) or 60-day windows (c); and blue color indicates either stepping down strategy (a,b) or 80 day windows (c). Finally, protection percentages can be seen in a, and social distance (SD) are shown ranging from 15-40% by the diameters of the circles in b and c.

Figure 3: Illustration of the influence of strategy (stepping down, intermittent, and constant) and of different time windows (40, 60, and 80 days) on the model results for active cases per day (a,i), total ICU beds per day (b,j), hospitalizations per day (c,k), accumulative deaths (d,l), accumulative cases (e,m), accumulative recuperated cases (f,n), as well as social distancing (SD) strategies (g,o), and protection percentage (h,p) for São Paulo. Mean SD was similar among all cases (26%) and protection was either 65 or 60%.

Figure 4: Illustration of the influence of different mean social distance (13, 26, and 52%) and different protection percentages (65, 60, and 55%) on the model results for active cases per day (a,i), total ICU beds per day (b,j), hospitalizations per day (c,k), accumulative deaths (d,l), accumulative cases (e,m), accumulative recuperated cases (f,n), as well as social distancing (SD) strategies (g,o), and protection percentage (h,p) for São Paulo. SD were either 13, 26 or 52% with a 30% protection percentage and protection was either 55, 60 or 65% with a 52% SD.

Table 1

| Coeffs | Fit_BR | Fit_SP | Lower B. | Upper B. |
|---|---|---|---|---|
| $\alpha$ | 0.024 | 0.018 | 0.01 | 0.12 |
| $\beta$ | 0.59 | 0.65 | 0.5 | 1.2 |
| $\gamma$ | 0.93 | 1.84 | 0.5 | 5 |
| $\delta$ | 0.07 | 0.11 | 0.07 | 0.5 |
| $\zeta$ | 0.25 | 0.24 | 0.2 | 0.33 |
| $\varepsilon$ | 0.10 | 0.11 | 0.05 | 0.14 |
| $m$ | 0.94 | 0.95 | 0.65 | 0.99 |
| $c$ | 0.36 | 0.35 | $E0/2$ | $2E0$ |
| $f$ | 0.49 | 0.50 | $I0/2$ | $2I0$ |
| $E0$ | 273 | 38 | $H0/2$ | $2H0$ |
| $I0$ | 195 | 18 | $C0/2$ | $2C0$ |
| $H0$ | 205 | 22 | $Re0/2$ | $2Re0$ |
| $C0$ | 0 | 0 | $D0/2$ | $2D0$ |
| $Re0$ | 0 | 0 | | |
| $D0$ | 0 | 0 | | |
| FactorC | 25.9 | 25.9 | | |
| FactorD | 1.9 | 1.7 | | |
| Latent | 1.1 | 0.5 | | |
| Infectious | 14.4 | 9.2 | | |
| Hospitalized | 4.0 | 4.1 | | |
| Critical | 10.0 | 9.2 | | |

Figure 1

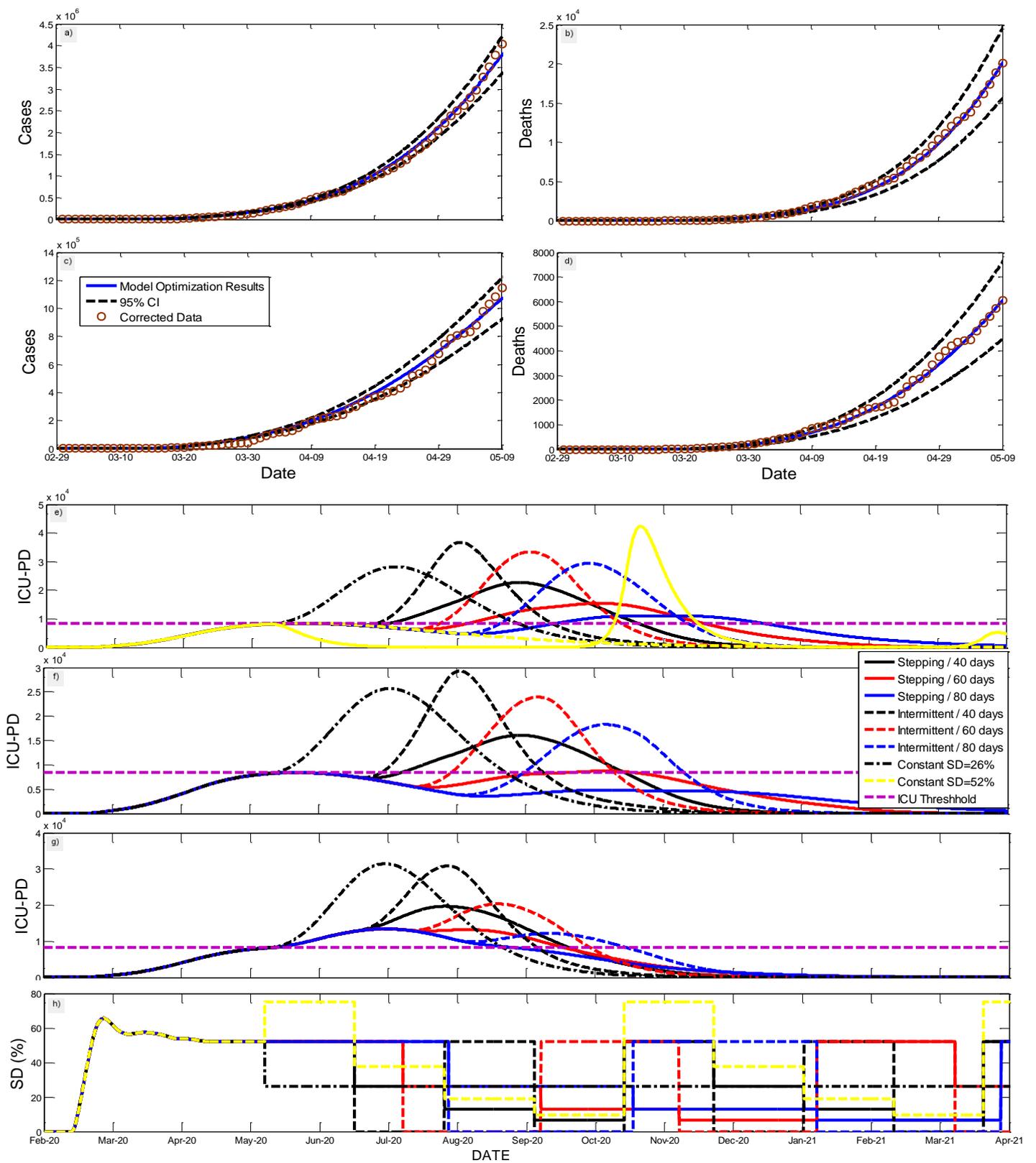

Figure 2

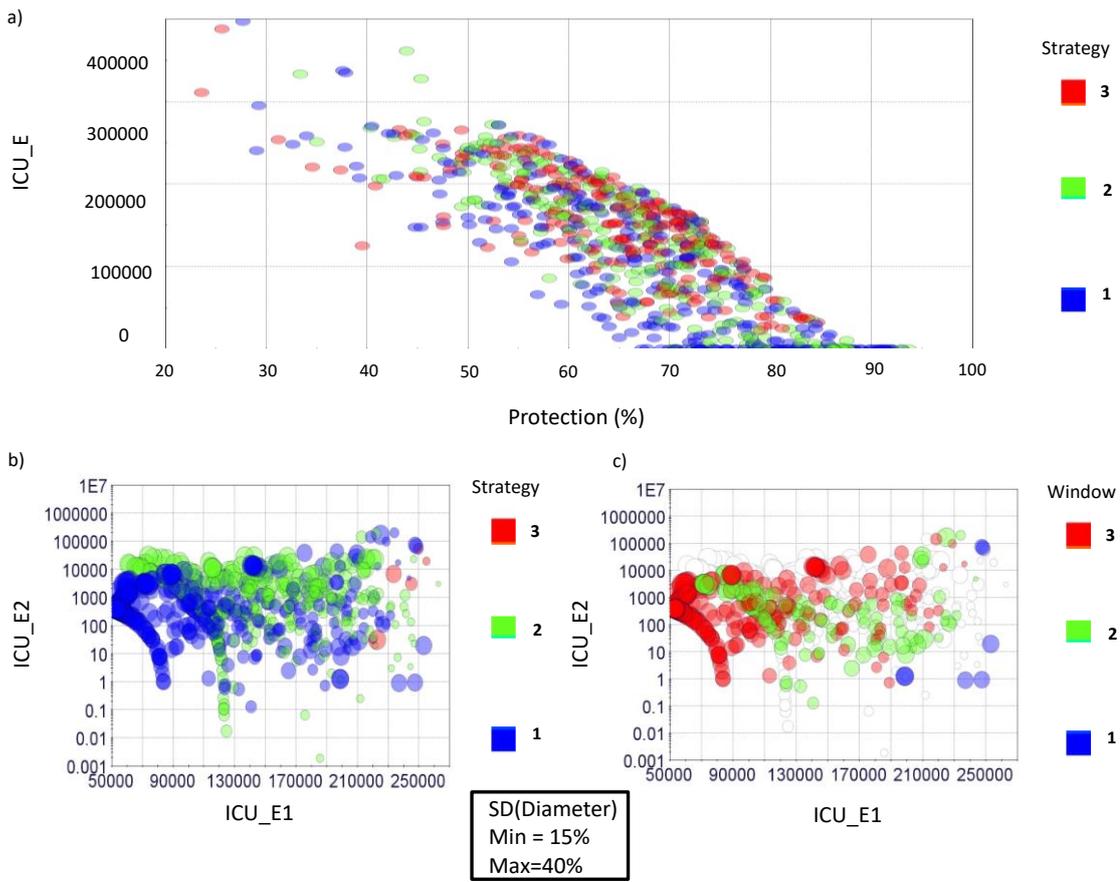

Figure 3

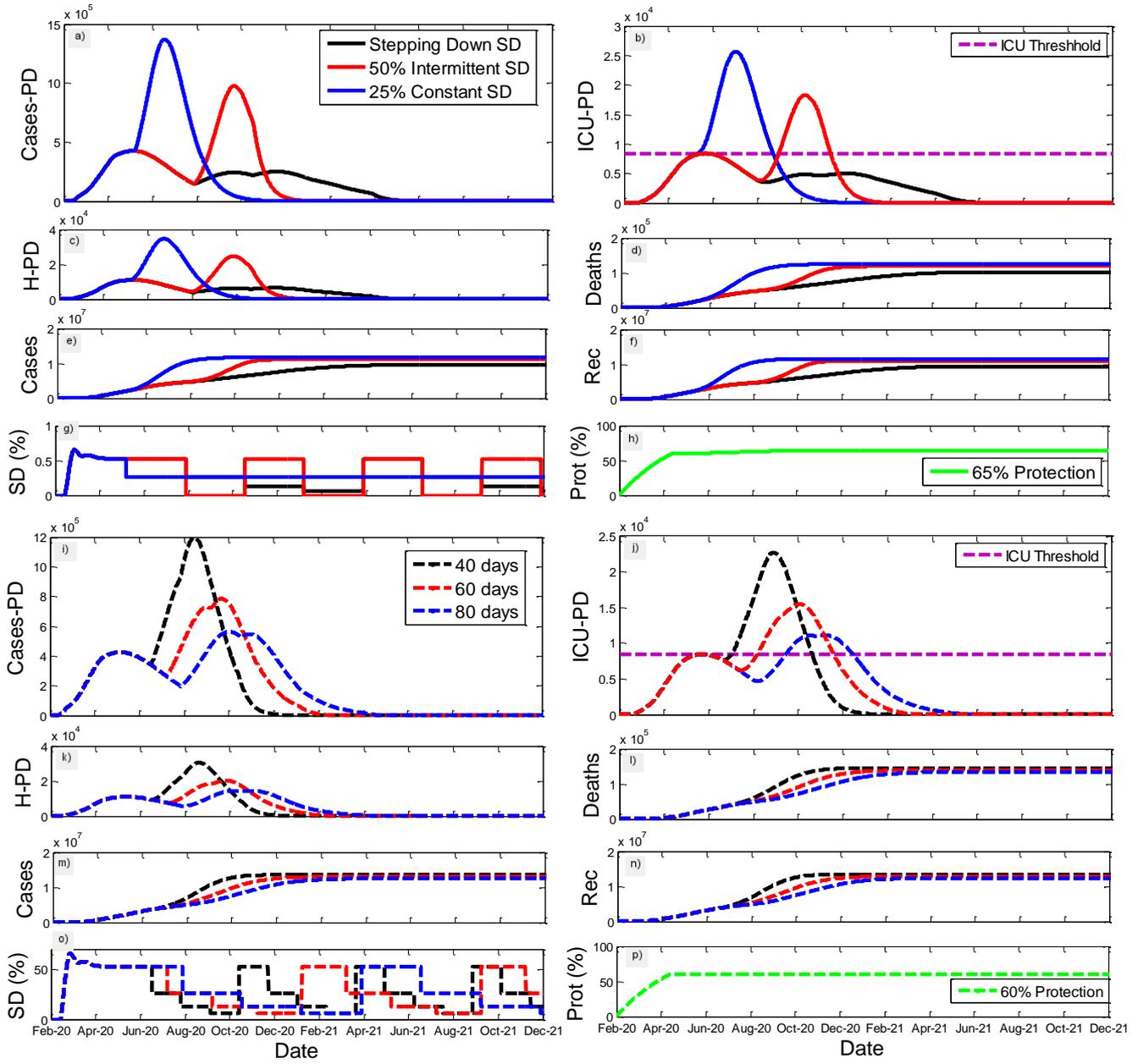

Figure 4

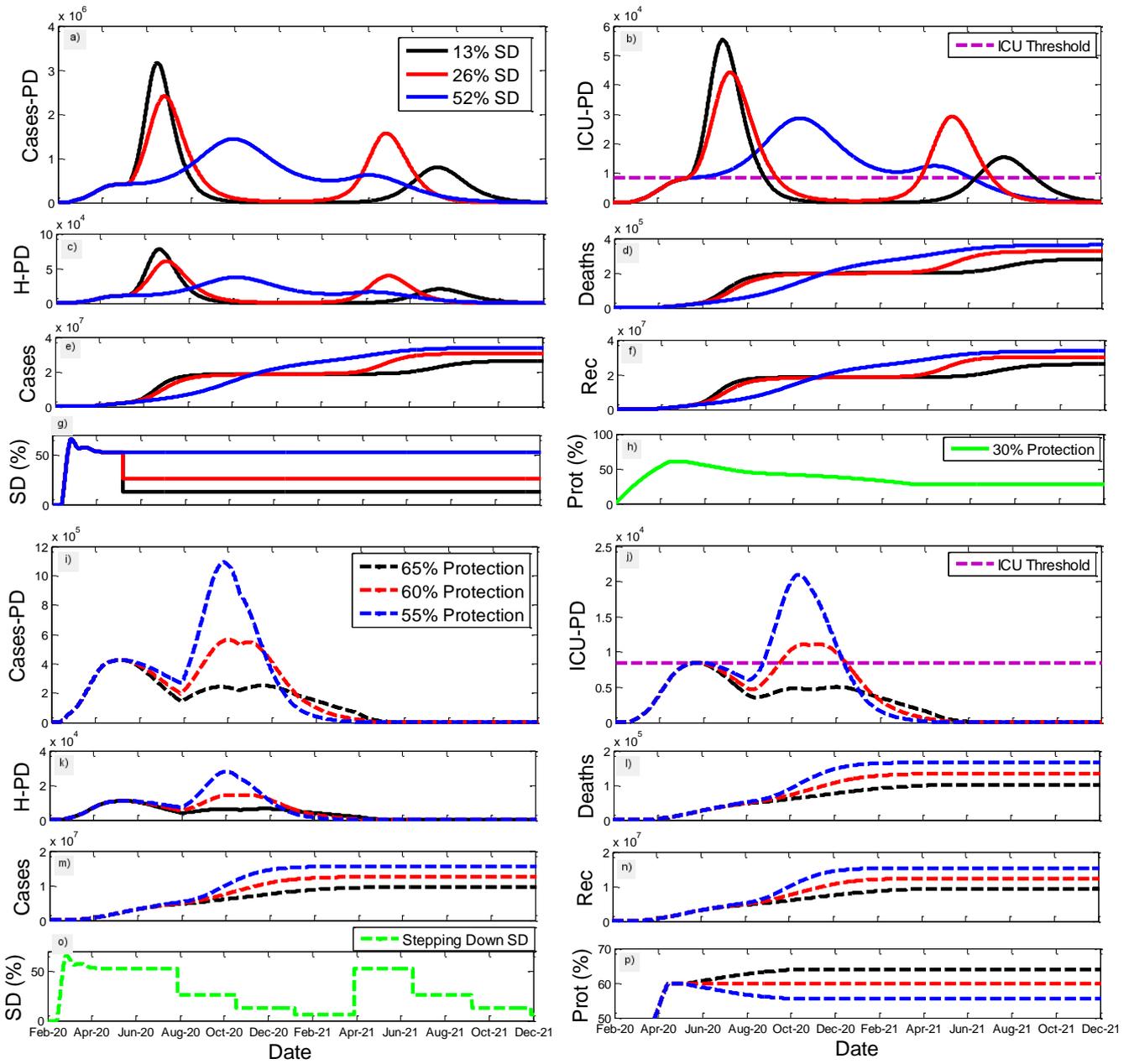